\documentclass{emulateapj}
\pdfoutput=1
\usepackage{apjfonts}

\usepackage[latin1]{inputenc}

\shorttitle{Broken profiles from cosmological simulations}
\shortauthors{Martínez-Serrano et al.}

\begin{document}
\title{Disk galaxies with broken luminosity profiles from cosmological simulations}

\author{F.J. Martínez-Serrano, A. Serna, M. Domenéch-Moral}
\affil{Depto. de Física y A.C., Universidad Miguel Hernández,	
E-03202 Elche, Alicante, Spain}

\and

\author{R. Domínguez-Tenreiro}
\affil{Depto. de Física Teórica, Universidad Autónoma de Madrid,
E-28049 Cantoblanco, Madrid, Spain}

\email{franjesus@umh.es}

\begin{abstract}

We present SPH cosmological simulations of the formation of three disk galaxies
with a detailed treatment of chemical evolution and cooling. The resulting
galaxies have properties compatible with observations: relatively high
disk-to-total ratios, thin stellar disks and good agreement with the
Tully-Fisher and the luminosity-size relations. They present a break in the
luminosity profile at $3.0 \pm 0.5$ disk scale lengths, while showing an
exponential mass profile without any apparent breaks, in line with recent
observational results. Since the stellar mass profile is exponential, only
differences in the stellar populations can be the cause of the luminosity
break. Although we find a cutoff for the star formation rate imposed by a
density threshold in our star formation model, it does not coincide with the
luminosity break and is located at $4.3 \pm 0.4$ disk scale lengths, with star
formation going on between both radii. The color profiles and the age profiles
are ``U-shaped'', with the minimum for both profiles located approximately at
the break radius. The SFR to stellar mass ratio increases until the break,
explaining the coincidence of the break with the minimum of the age profile.
Beyond the break we find a steep decline in the gas density and, consequently,
a decline in the SFR and redder colors. We show that most stars (64-78\%) in
the outer disk originate in the inner disk and afterwards migrate there. Such
stellar migrations are likely the main origin of the U-shaped age profile and,
therefore, of the luminosity break.

\end{abstract}


\section{Introduction}

There is a growing body of observations that show that the classical picture of
spiral galaxies with a purely exponential disk does not match most galaxies
found in the local \citep{2002A&A...392..807P, 2006A&A...454..759P,
2007MNRAS.378..594P, 2007A&A...466..883V, 2008ApJ...683L.103B} or distant
universe \citep{2004A&A...427L..17P, 2005ApJ...630L..17T, 2008ApJ...684.1026A}.
Indeed, galactic disks with a single exponential surface brightness profile
\citep[Type-I, in the classification of][]{1970ApJ...160..811F} are relatively
rare ($\lesssim 15$\%). Most profiles are better fitted by two exponentials
separated at a well defined break radius, with the outer exponential being
either downbending (Type-II, the majority) or upbending \citep[Type-III, first
discovered by][]{2005ApJ...626L..81E}. The profiles with an upbending break
could be explained \citep[][PT06 hereafter]{2006A&A...454..759P} either from a
disturbed morphology due to interactions with neighboring galaxies or, in some
cases, by the presence of an $R^{1/4}$ bulge component that rises over the
exponential disk in the outer region. The origin of Type-II profiles, more
frequent in late-type galaxies, is instead much more poorly understood. 

Several authors have investigated how galaxy disks develop a downbending break,
mainly through non-cosmological N-body simulations of isolated objects
\citep{2006ApJ...645..209D, 2007ApJ...670..237B, 2008ApJ...684L..79R,
2008ApJ...675L..65R, 2008MNRAS.386.1821F}, but also from cosmological
simulations \citep{2003ApJ...591..499A,2007MNRAS.374.1479G}. All these works
agree in explaining the observed break in the surface brightness profile as the
result of an intrinsic break in the mass profile for the stellar component. In
the most widely considered scenario, the break in the stellar mass profile
develops as a consequence of a cutoff in the star formation
\citep{1994ApJ...435L.121E}. In such a scenario the outer exponential disk
would be formed by stars that migrate from the inner disk towards the regions
beyond the star formation cutoff, prompted by secular instabilities such as
spiral arms \citep{2008ApJ...684L..79R, 2008ApJ...675L..65R} or clump
disruptions \citep{2007ApJ...670..237B}. An alternative model has been
suggested by \citet{2008MNRAS.386.1821F} after simulations starting from an
already formed single exponential disk. In their simulations the inner disk
forms when the bulge draws mass from the inner regions, thus making the inner
disk profile shallower while the outer region stays almost unaltered and fixed
by the initial conditions.

All the above models fail however to match recent observational evidence
provided by \citet{2008ApJ...683L.103B}\footnote{A work by
\citet{2009arXiv0905.4579S} appeared about the time this paper was submitted
showing that, in agreement with our work, broken luminosity profiles can appear
in simulated galaxies with exponential mass profiles.}. These latter authors
have analyzed a sample of late-spirals and found that galaxies with a Type-II
surface brightness profile generally have a "U-shaped" color profile with its
minimum nearly located at the luminosity break radius. When this color profile
was taken into account to calculate the stellar surface mass density, they
obtained almost purely exponential mass profiles. Consequently, the observed
luminosity break in Type-II galaxies does not seem to be necessarily related to
an intrinsic break in the stellar mass profile. A downbending of the luminosity
profile could also be caused by differences between the stellar populations of
the inner and outer disk, even with no breaks in the mass profile at all.
However, stellar counts of RGB stars in the outskirts of M33
\citep{2007iuse.book..239F} or NGC 4244 \citep{2007ApJ...667L..49D} indicate
that this may not always be the case.

The aim of this letter is to study whether numerical simulations can naturally
predict the above observed features: downbending broken luminosity profiles in
galactic disks with a purely exponential mass distribution. We will use
cosmological simulations where galaxies form after the collapse of primordial
instabilities and where their final structure arises from a hierarchical
sequence of mass aggregation and from the interaction with their environments.

\section{The simulations}

The N-Body + SPH simulations have been performed using an OpenMP parallel
version of DEVA code \citep{2003ApJ...597..878S} and the methods for star
formation, chemical feedback and abundance-dependent cooling described in
\citet{2008MNRAS.388...39M}. In this code, particular attention has been paid
that the conservation laws hold as accurately as possible. Star formation is
implemented through a Kennincutt-Schmidt-like law with a density threshold of
$\rho_{\rm thres}=6.75\times10^{-26}$ g/cm$^3$ and a star formation efficiency
of $c_* = 0.1$, that implicitly account for energy feedback\footnote{We use
this simple model for subresolution physics because our aim in this paper is to
test the minimal conditions for the formation of realistic disks.}. Each
simulation is a cosmological \textit{zoom-in} that includes high-resolution gas
and dark matter for the flow converging region that generates the main object.
The rest of the simulation box is sampled by low-resolution dark matter
particles that account for tidal forces over the flow converging region. As a
first step we consider a concordance cosmological model and generate three
high-resolution initial conditions in a box of 10 Mpc per
side\footnote{Although this box side implies a lack of very massive objects and
environments, it has little effect on the internal properties of the haloes
\citep{2006MNRAS.370..691P}.} After degrading these conditions we perform the
corresponding full box simulations at lower resolution. In each of these three
simulations, we selected one object (the most massive one with a prominent gas
disk at $z=0$ not clearly disrupted from a very recent major merger) and
traced back the particles inside its virial radius until the initial redshift
$z_{init}$. We then computed the convex hull \citep{barber96quickhull}
enclosing these particles at $z_{init}$ and substituted all the particles
inside the convex hull by their high resolution counterparts. Gas particles
outside the hull are eliminated and their masses added to the low-resolution
dark matter component, thus obtaining the initial conditions for each
simulation. The galaxies obtained in this way have a full history of mergers
and accretion in a cosmological context, without any assumptions made for their
initial conditions beyond the cosmology and the initial conditions generator
used \citep{2008ApJS..178..179P}. The mass resolution for baryonic particles
used in each simulation is specified in Table \ref{tbl1}.

\section{Disk object properties}

\subsection{Consistency with observational data}

\begin{figure}
\plotone{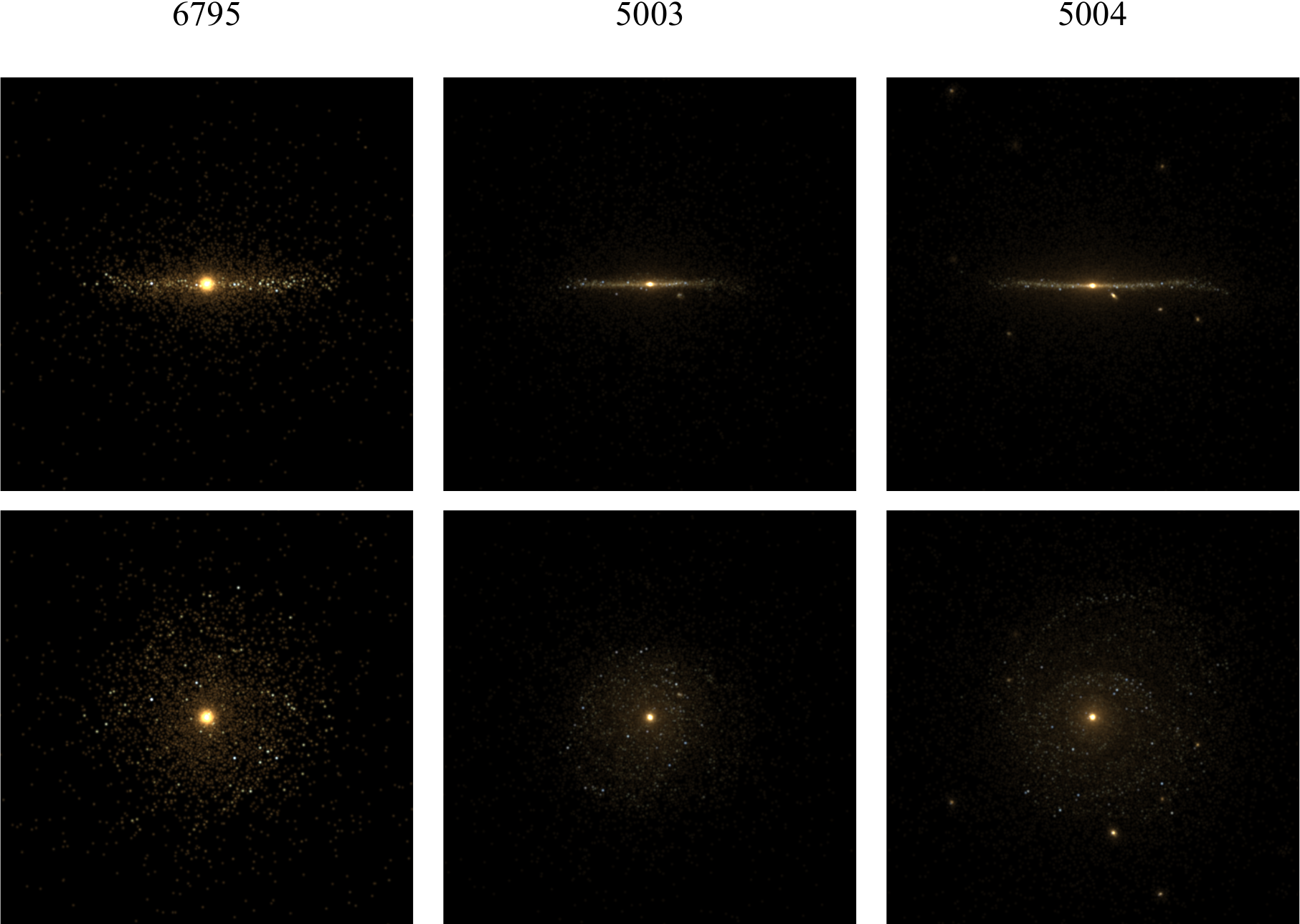}

\caption{Face-on and edge-on synthetic images obtained using \citet{2003MNRAS.344.1000B} models. All images are 50 kpc side.
\label{fig1}}
\end{figure}

Object properties are summarized in Table \ref{tbl1}, where we can see a
remarkable consistency with observational data (Fig. \ref{fig1} displays their
face-on and edge-on images). They span a range of stellar masses from 1.71 to
3.86 $\times$ 10$^{10}$ M$_\odot$. Their luminosity profiles have been obtained
and fitted by following the procedures described in $\S$\ref{sec:profile} below
(see also the notes of Table 1). A first remarkable feature of our simulations
is that we obtain objects with bulge sizes comparable to those recently
obtained by \citet{2008arXiv0812.0379G}. Indeed, two out of the three simulated
objects have D/T ratios implying rather small bulges, consistent with those
observed for late-type spirals \citep[\textit{e.g.}][]{2007ApJ...665.1104B,
2009MNRAS.393.1531G}. This is an interesting property because, although Type-II
breaks appear in all spiral types, they are more abundant in late Hubble types
\citep[PT06,][]{2008AJ....135...20E}. Note also that the disks are rather thin
with vertical scales of $\sim$ 0.4 -- 0.6 kpc for the thin disk and $\sim$
1.5-- 2.2 kpc for the thick disk. We also display in Table \ref{tbl1} the
I-band absolute magnitude and the gas rotation velocity of each object. They
are found to be in good agreement with the observed Tully-Fisher relation
\citep[\textit{e.g.}][see Fig. 6 of the latter reference]{1997AJ....113...22G,
2007ApJS..172..599S}. In the same way, through Sérsic fits to the total light
distribution we have determined the \textit{r}-band absolute Sérsic magnitude
M$_{r,s}$ and the half-light radius $r_{50,s}$ (see Table \ref{tbl1}). The
obtained values are again in good agreement with both the luminosity-size and
stellar mass-size relations for disk galaxies given by
\citet{2003MNRAS.343..978S} (see their Figs. 6 and 11, respectively).

\subsection{Broken face-on profile}\label{sec:profile}

\begin{figure}
\plotone{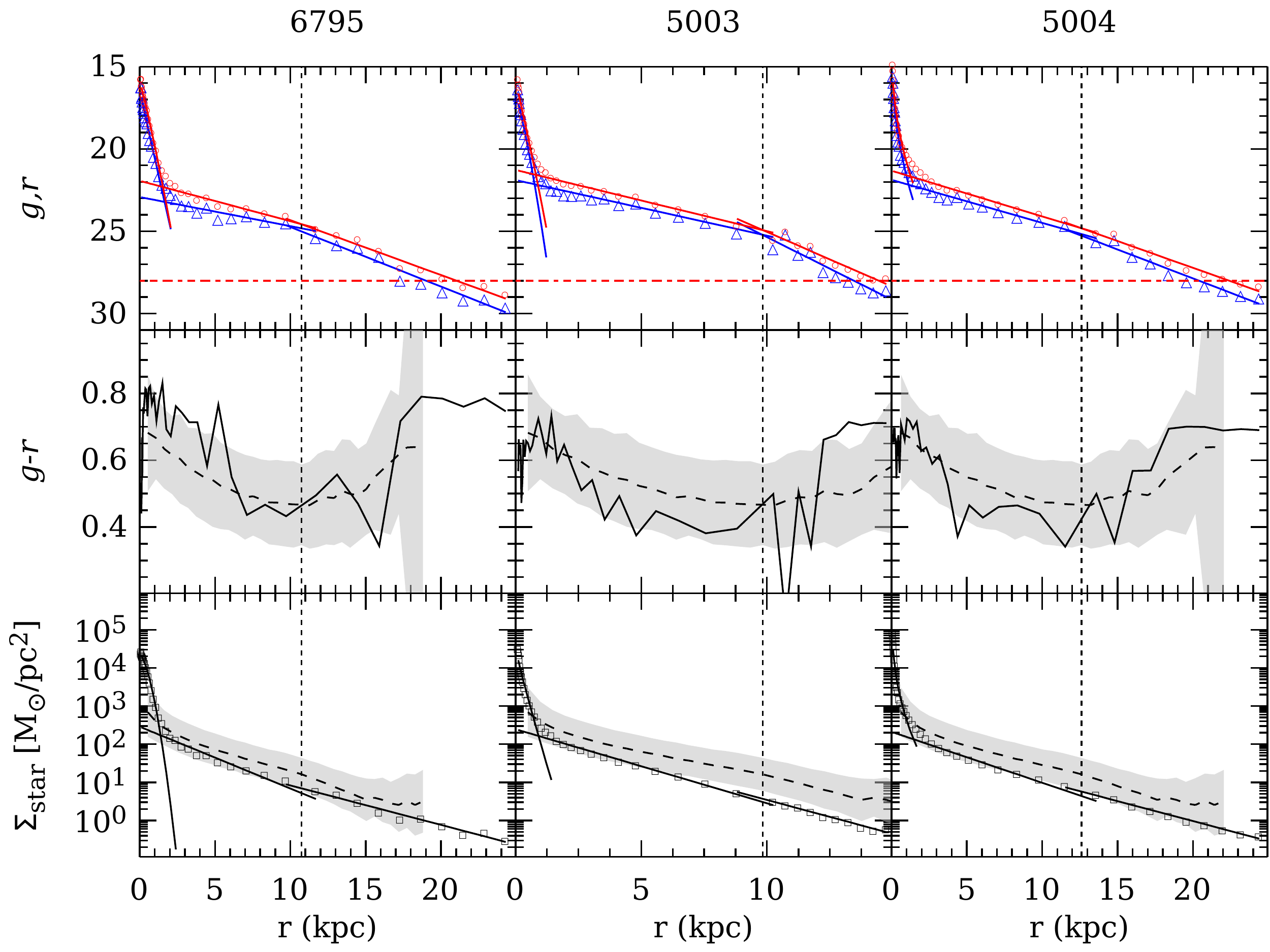}

\caption{Observational properties of the galaxies presented in this letter. The
top panels show the face-on luminosity profile for each galaxy in the
\textit{r} (circles) and \textit{g} (triangles) bands, together with fits to
the bulge and both disk components. Units are mag$/\Box^{\prime\prime}$. The
break radius for each galaxy separating the inner and outer exponential
components is shown as a dashed vertical line and an approximate critical
surface brightness of $\mu_{\rm lim} = 28.0\ r$-mag$/\Box^{\prime\prime}$. The
middle row shows the \textit{g}-\textit{r} color profile \citep[compare
with][]{2008ApJ...683L.103B}. The bottom row shows the mass profile which
unlike the luminosity profile is either exponential anti-trucated. For
reference, we have overplotted the \citet{2008ApJ...683L.103B} data for the
color and mass plots as dashed lines with the scatter of the observational data
shown as a gray shaded area.  \label{fig2}}

\end{figure}

In the first row of Fig. \ref{fig2} we plot the azimuthally averaged surface
luminosity profiles in both the \textit{g} and \textit{r} bands for the three
simulated galaxies, along with an approximate limit surface luminosity
$\mu_{\rm lim}$ beyond which SDSS observations such as those of PT06 become
background-dominated. These profiles were obtained by computing the luminosity
of each stellar particle, essentially a single stellar population (SSP) with
known age and metallicity, and interpolating in the \citet{2003MNRAS.344.1000B}
tables. We note that all these profiles can indeed be considered as broken
exponentials. The slope in the outer disk region is in all cases steeper than
in the inner disk and, hence, the three simulated objects can be classified as
Type-II galaxies. We have then fitted each profile by means of a
Levenberg-Marquardt algorithm based on a double exponential + bulge model. In
order to reduce noise, the fits were computed by using the integrated
luminosity profiles instead of the surface luminosity profiles. The resulting
bulge and disk scale lengths are given in Table \ref{tbl1}, as well as the
break radius (defined by the intersection of the inner and outer exponentials)
of each galaxy. We find that the average ratio between the break radius and the
inner exponential disk scale is $r_b/r_s=3.0 \pm 0.5$, in agreement with the
value found by PT06.

We also note that the \textit{g} and \textit{r} profiles have different slopes,
both before and after the break radius. This results in characteristic
``U-shaped'' \textit{g} - \textit{r} profiles such as those depicted in the
second row of Fig. \ref{fig2}. These color profiles are very similar to those
observed in many spiral galaxies with broken profiles
\citep{2008ApJ...683L.103B}. The stellar mass profiles however do not show any
downbending trend. On the contrary, as row 3 of Fig. \ref{fig2} shows, mass
profiles can be considered as Type I (no break) in all the simulated galaxies,
or even with a very slightly upbending trend (Type III). This result agrees
with the findings of \citet{2008ApJ...683L.103B} who used the M/L ratio
provided by \citet{2003ApJS..149..289B} to derive a mass profile from
observational data. In any case, outside the luminosity break we do not find
any lack of stellar mass respect to a single exponential decline. 

\begin{figure}
\plotone{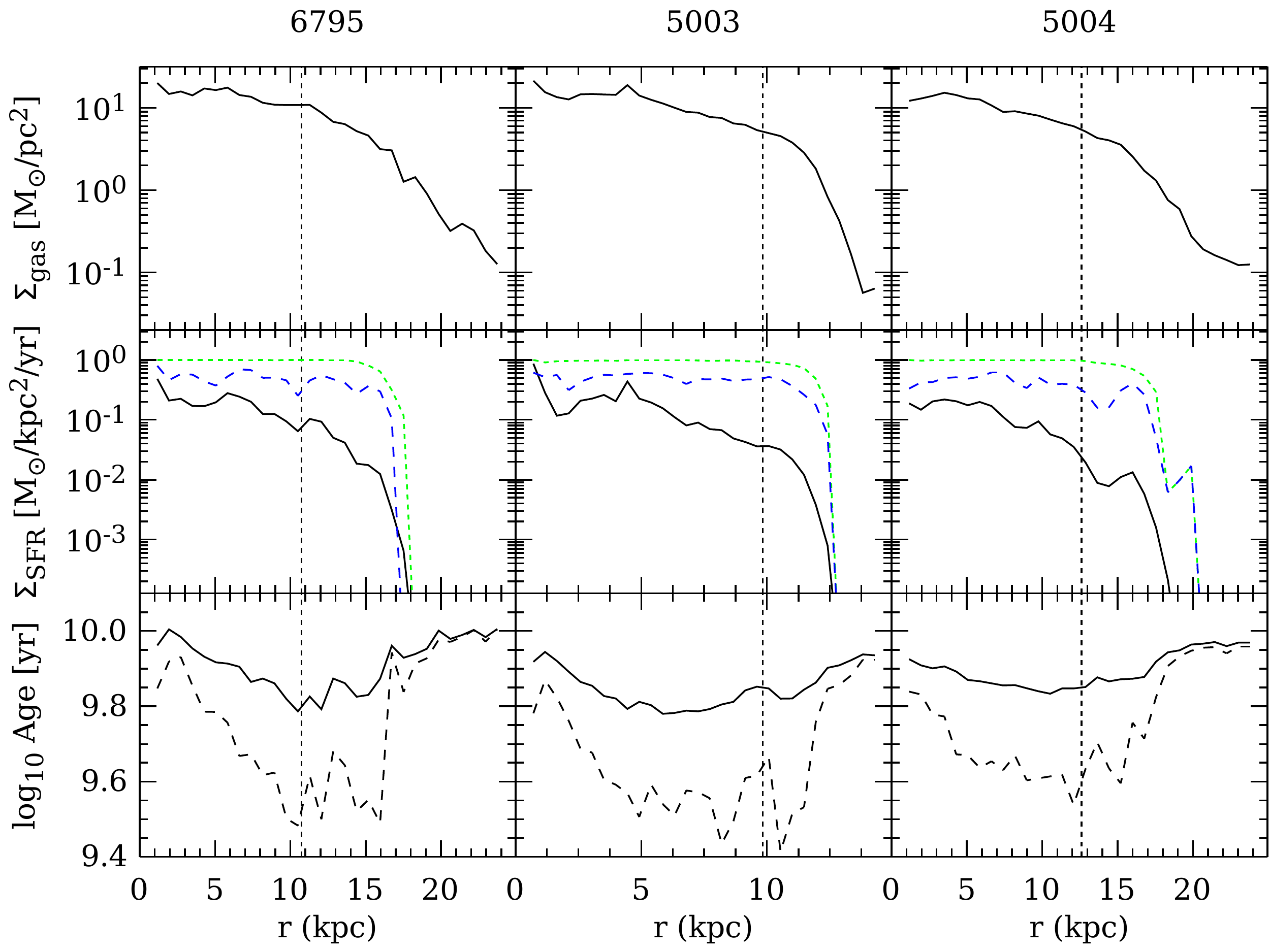}

\caption{Dynamical properties of the galaxies presented in this work as a
function of radius. The top panels show the gas column-density profile. As in
Figure \ref{fig2}, the dashed vertical line shows the computed break radius for
each galaxy in the \textit{r} band. The middle row shows the instantaneous star
formation rate derived from the Schmidt law implemented in the code as a
continuous line, with the fraction of gas particles able to form stars
according to the $\rho_{\rm thres}$ limit and both the $\rho_{\rm thres}$ and
$\nabla \cdot \mathbf{v}<0$ criterion shown as a dotted and dashed lines
respectively. The bottom row shows the mean stellar age weighted by stellar
mass (continuous line) and \textit{r}-band luminosity (dashed line).
\label{fig3}}

\end{figure}

In order to progress in understanding the nature of the above downbending
luminosity profiles, we have analyzed the possible connection between the break
radius and the cutoff radius for star formation. In the first row of Fig.
\ref{fig3} we represent the surface gas density for each galaxy. For all of
them the profile shows a plateau that extends from the center of the galaxy
with a shallow slope until the break radius, marked with a dotted vertical
line. Given the Schmidt-Kennicutt star formation recipe implemented in our code
\citep{2008MNRAS.388...39M}, the star formation rate (SFR) is expected to
follow the gas density at a power 1.5 and increase towards the centre. Indeed,
in the second row we depict the instantaneous SFR at $z=0$, which follows a
similar trend, with a steeper slope, especially for $r>r_b$. On the same row it
is also represented the fraction of gas particles above the density threshold
for star formation, and the fraction of gas particles eligible to form stars
according to both the density criterion and the convergent flow ($\nabla \cdot
\mathbf{v}<0$) criterion. As it can be seen, the density criterion effectively
imposes a cutoff radius for the star formation, while the convergent flow
criterion does not impose any significant restriction on the amount of gas
particles able to form stars at a given radius, appearing just as some
oscillations due to the spiral arm structure of the objects that compresses and
rarefies the gas. The position of the $\rho_{\rm thres}$ cutoff is always well
outside the break radius, implying that our choice of $\rho_{\rm thres}$ does
not directly impose the position of the luminosity break, but it rather
appears well within the zone where star formation is allowed. Note also that
the cutoff position coincides approximately with a surface gas density (first
row of Fig. \ref{fig3}) of $\sim 1$ M$_\odot$/pc$^2$, while the density at the
break is $\sim 10$ M$_\odot$/pc$^2$, coinciding with the value given by
\citet{2001ApJ...555..301M} for the threshold of star formation.

In the works of \citet{2008ApJ...675L..65R, 2008ApJ...684L..79R}, a break in
the mass profile appears in the same position where a significant drop in the
SFR happens, with the stars beyond the break being transported there by
migrations from the inner parts of the galaxy. Our galaxies instead do not show
such a steep break for the SFR and consistently do not present a break on the
stellar mass profile, so that the only interpretation for this fact is a
difference in the stellar populations in each region. Indeed, as row 3 of Fig.
\ref{fig3} shows, there is a ``U-shape'' profile for the mean stellar age, very
similar in shape to the color profile. The break on the profile appears to be
at the minimum of the stellar age profile for galaxies 6795 and 5004, while
5003 presents a larger minimum. The coincidence between the break radius and
the minimum of the age profile was already noted by
\citet{2008ApJ...675L..65R}. Given that young stellar populations are brighter
than old ones, the age decrease until the break radius makes the light profile
shallower than the mass profile, while the increase after the break radius
makes it steeper. This fact alone would explain the luminosity break. It also
readily accounts for the color profile, since younger populations are bluer. 

\section{Origin of the stars in the outer disk and discussion}

Our interpretation of the appearance of a break in the luminosity profile is
related with the interplay between the slope of the stellar mass profile,
exponential without any significant change, and that of the gas profile, which
is shallower in the inner part and drops rather abruptly beyond the break
radius. The shallow inner profile of the gas can be attributed to angular
momentum transfer from the inner particles towards the outer parts, possibly
due to spiral arm instabilities \citep{2007MNRAS.375...53K} or viscosity
effects. The steeper profile of the inner stellar disk is likely due to the
inside-out grow mode of our disks: the smaller the radius the longer star
formation has been going on there. Going from the centre towards the break
radius, the ratio of gas mass to stellar mass, and consequently, the ratio of
SFR to stellar mass increases. This explains the coincidence of the break
radius with the minimum of mean stellar age. Once we reach the break radius,
there is a change in the slope of the gas density profile that can be
interpreted as an ending of the gas disk. This is likely to be due to
cosmological effects such as inflow of gas with angular momentum not aligned
with the disk rotation. 

\begin{figure}
\plotone{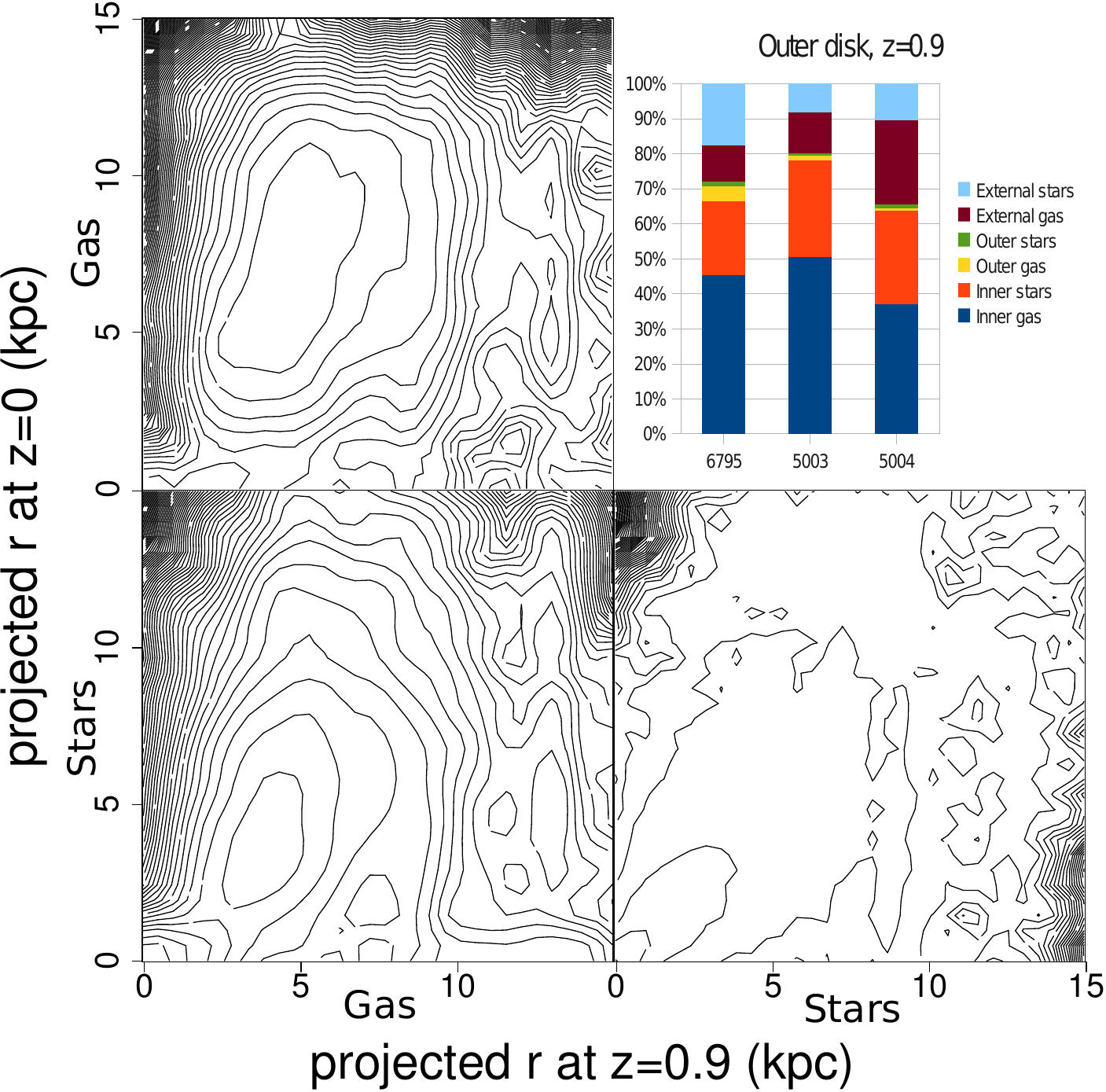}

\caption{Top-right panel: Location at $z=0.9$ of the outer disk stars at $z=0$. As an example, the other panels contain contour plots of the evolution of the radial position of particles from $z=0.9$ to $z=0$.  \label{fig4}}

\end{figure}

Independently of the mechanism that leads to a cutoff of the gaseous disk at
large radii, there is a transition zone between the break and the cutoff radius
where star formation is still possible, but the slope of the SFR is steeper
than that of the stellar disk. There must be a mechanism that brings stars to
this area, since star formation alone cannot account for all the stars present
there. We have analysed the origin of the stars present in the outer disk
($r>r_b$) at $z=0$ by tracing them back in time. We divide the volume of each
simulation into three regions at $z=0.9$. We assume a disk height of 2 kpc and
define the inner disk as a cylinder with $r<r_b$ and the outer disk as a volume
with $r_b<r<r_c$. The rest of the volume is considered to be external. In the
top-right panel of Fig.\ref{fig4} we show the fraction of the outer disk stars
that belong to each of these volumes, separated also into gas and stars (likely
old and moderately old stars at $z=0$, respectively). As it can be seen, most
stars (64--78\%) in the outer disk were residing in the inner disk at $z=0.9$,
either as gas (the majority), or stars. This indicates that stellar migrations
play a relevant role in the build up of the outer disk, as
\citet{2008ApJ...684L..79R} pointed out. Few of the particles (2--6\%) were
already present in the outer disk, this is expected since the disk size is
significantly smaller at that time, and by no means should be taken as an
indication that the present SFR in the outer disk has such a little weight. The
rest (20--34\%) were located outside of the galaxy either in smaller objects
that merged in, either as diffuse gas. The fact that most of the stars in the
outer disk originate in the inner disk means that the star formation taking
place in the outer disk is sub-dominant when compared with the stars that got
transported there by secular processes, possibly driven by spiral arms, or got
scattered there in a merger process. In the rest of panels of Fig.\ref{fig4} we
show for one of the galaxies the radial distances of different types of
particles at $z=0.9$ versus their corresponding distances at $z=0$. The
bottom-right panel displays stellar particles born before $z=0.9$ (old-star
migrations), while the bottom-left shows particles that become stellar after
$z=0.9$ (younger star progenitor migrations). Finally, in the top-left panel we
give results for gas particles. We can clearly see that old stellar particles
have migrated to the outer region. The trend for stars born after $z=0.9$ and
with $r < $3 kpc at $z= 0.9$ is even clearer, although these are a minority.
Gaseous particles also migrate. We then conclude that such migrations are
likely the main origin of the U-shaped age profile and, therefore, of the
luminosity break.

\acknowledgements{We are indebted to Ignacio Trujillo for useful discussions
and suggestions and Judit Bakos for kindly providing us their observational
data. We also thank the anonymous referees for their helpful comments and
suggestions. This work was partially supported by the Ministerio de Ciencia e
Innovación, Spain, through grants AYA2006-15492-C03-01 and AYA2006-15492-C03-02
from the PNAyA, and grant CSD-2007-00050 from the Consolider Ingenio-2010
program. It was also supported by the Madrid IV PRICIT program through the
ASTROCAM Astrophysics network (S-0505/ESP-0237). We thank the Centro de
Computación Científica (UAM, Spain) and the Barcelona Supercomputing Center for
computing facilities and support.} 

\bibliographystyle{hapj}

\clearpage

\begin{table}
\begin{center}
\caption{Objects analyzed\label{tbl1}}
\begin{tabular}{lccccccccccccccc}
\tableline\tableline

Galaxy & m$_{\rm bar}$\tablenotemark{a} & M$_{\rm star}$\tablenotemark{b} & D/T\tablenotemark{c} & $r_e$\tablenotemark{c} & $r_b$\tablenotemark{c} & M$_\mathrm{I}-5\log(h)$\tablenotemark{d} & $h_{z1}$\tablenotemark{e} & M$_{r,s}$\tablenotemark{f}  \\ &  $\epsilon$ & M$_{\rm gas}$\tablenotemark{b} & $r_s$\tablenotemark{c} & $n$\tablenotemark{c} & $r_c$\tablenotemark{c} & V$_\mathrm{rot}$\tablenotemark{d} & $h_{z2}$\tablenotemark{e} & $r_{50,s}$\tablenotemark{f} \\
\tableline
	                & [M$_\odot$] & [$10^{10}$M$_\odot$] & & [kpc] & [kpc]             &                       &     [kpc]           &  \\     & [kpc]                  &  [$10^{9}$M$_\odot$] &  [kpc]      & & [kpc]        & km/s    & [kpc]  & [kpc]     \\

\tableline
6795 & 2.92$\times10^6$ & 3.86 & 0.39 & 0.45 & 10.75 & -20.75 & 0.50 & -20.82 \\ & 0.55 & 9.59 & 4.42 & 0.87 & 18.1 & 203.6 & 1.53 & 2.32 \\
\tableline                                                                  
5003 & 3.82$\times10^5$ & 1.71 & 0.67 & 0.26 & 9.83 &  -20.14 & 0.42 & -20.23 \\  & 0.40  & 4.15 & 2.92 & 0.91 & 12.1 & 133.2 & 1.99 & 1.22 \\
\tableline                                                                  
5004 & 3.82$\times10^5$ & 3.46 & 0.64 & 0.31 & 12.60 &  -20.78 & 0.59 & -20.88 \\ & 0.40 & 7.25 & 3.92 & 2.33 & 20.2 & 162.2 & 2.14 & 2.66 \\
\tableline
\end{tabular}

\tablenotetext{a}{Mass of a single baryonic particle (gas or star).}
\tablenotetext{b}{This value was computed using the particles inside ten
times the disk radius $r_s$ for each object.}
\tablenotetext{c}{$r_s$ is the inner disk scale length or e-folding length.
$r_e$ is the effective radius of the bulge. $r_b$ is the break radius. $r_c$ is
the cutoff radius. $n$ is the S\'ersic parameter for the bulge. The fits are for
the \textit{r}-band luminosity profile, obtained using the models of
\citet{2003MNRAS.344.1000B}.}
\tablenotetext{d}{Total I-band luminosity and gas rotation speed.}
\tablenotetext{e}{Vertical mass profile scaleheights for the thin and thick
stellar disks. Obtained using particles with $r_e<r<r_b$.}
\tablenotetext{f}{S\'ersic fits to the total light distribution. Given are
S\'ersic's half-light radius and total \textit{r}-band luminosity.}

\end{center}
\end{table}

\end{document}